\documentclass[letterpaper, 10pt, conference]{ieeeconf}                              
\usepackage{cite}
\usepackage{amsmath,amssymb,amsfonts}

\usepackage{algorithm}
\usepackage{algorithmic}
\usepackage{bbold}
\usepackage{graphicx}
\usepackage{textcomp}
\usepackage{braket}
\usepackage{xcolor}
\usepackage{lipsum}
\usepackage{epsfig}
\usepackage{tikz}
\usepackage{float}
\usetikzlibrary{shapes.geometric, arrows, quantikz, backgrounds, patterns, shapes.symbols, decorations.pathmorphing}
\def\BibTeX{{\rm B\kern-.05em{\sc i\kern-.025em b}\kern-.08em
    T\kern-.1667em\lower.7ex\hbox{E}\kern-.125emX}}

\title{\LARGE \bf Formulation of the Electric Vehicle Charging and Routing Problem for a Hybrid Quantum--Classical Search Space Reduction Heuristic}

\author{Mikel Garcia de Andoin$^{1,2,3,*}$, Alberto Bottarelli$^{4,5,*}$, Sebastian Schmitt$^6$,\\Izaskun Oregi$^{2}$, Philipp Hauke$^{4,5}$, and Mikel Sanz$^{1,3,7,8}$\\
$^1$ Department of Physical Chemistry, University of the Basque Country UPV/EHU, 48940 Leioa, Spain\\
$^2$ TECNALIA, Basque Research and Technology Alliance (BRTA), 48160 Derio, Spain\\
$^3$ EHU Quantum Center, University of the Basque Country UPV/EHU, 48940 Leioa, Spain\\
$^4$ Pitaevskii BEC Center, CNR-INO and Dipartimento di Fisica, Università di Trento, I-38123 Trento, Italy\\
$^5$ INFN-TIFPA, Trento Institute for Fundamental Physics and Applications, Trento, Italy\\
$^6$ Honda Research Institute Europe GmbH, 63073 Offenbach, Germany\\
$^7$ IKERBASQUE, Basque Foundation for Science, 48009 Bilbao, Spain\\
$^8$ Basque Center for Applied Mathematics BCAM, 48009 Bilbao, Spain\\
$^*$ Corresponding authors: mikel.garciadeandoin@ehu.eus, alberto.bottarelli@unitn.it\\
}

\begin{document}

\maketitle

\begin{abstract}
Combinatorial optimization problems have attracted much interest in the quantum computing community in the recent years as a potential testbed to showcase quantum advantage. In this paper, we show how to exploit multilevel carriers of quantum information---qudits---for the construction of algorithms for constrained quantum optimization. These systems have been recently introduced in the context of quantum optimization and they allow us to treat more general problems than the ones usually mapped into qubit systems. In particular, we propose a hybrid classical quantum heuristic strategy that allows us to sample constrained solutions while greatly reducing the search space of the problem, thus optimizing the use of fewer quantum resources. As an example, we focus on the Electric Vehicle Charging and Routing Problem (EVCRP). We translate the classical problem and map it into a quantum system, obtaining promising results on a toy example which shows the validity of our technique.  
\end{abstract}

\section{Introduction}
Optimization problems are ubiquitous in modern science and engineering, ranging from the design of logistic operations~\cite{Laurent2016} to the optimization of financial portfolios~\cite{Kolm2014}. For these problems to be non-trivial, a crucial requirement is the introduction of constraints into the solutions~\cite{Boyd2004}. Typically for classical industry-relevant optimization problems, the number of free parameters handled by classical optimization methods quickly meets the exponential computational threshold and thus they quickly become inefficient.

As a way out, two different strategies are possible: utilizing approximate algorithms\cite{hochba1997approximation} or turning to quantum computation~\cite{NielsenChuang2010}. While the first approach has been pursued for many decades, the latter field has become very active in particular in the last 10 years, with two main candidates showing promising results: Quantum Annealing\cite{hauke2020perspectives} and Quantum Approximate Optimization Algorithm (QAOA) \cite{kadowaki1998, rolando2012, farhi2014, passarelli2023}. Recently, researchers from the quantum computing community have been focusing on the developing of algorithms that make use of qudits (i.e., $d-$level systems) in order to tackle problems with tools that can be more versatile and have a larger information density than regular qubits.

In this paper, we give an example of the beneficial use of qudits by mapping the highly nontrivial Electric Vehicle Charging and Routing Problem (EVCRP) into the ground state computation of an interacting qudit system. This is an important optimization problem in the transportation industry that involves finding the most efficient way to charge electric vehicles while simultaneously determining their optimal routing. Due to the increasing popularity of electric vehicles and the need to develop sustainable transportation systems, this problem is becoming ever more relevant.  It is highly complex, with a large number of constraints that need to be considered, such as the limited range of electric vehicles and the availability of charging stations. The importance of solving the EVCRP lies in its potential to reduce the carbon footprint of transportation and make it more sustainable. By finding the most efficient way to charge electric vehicles and determine their routes, we can optimize the use of resources, reduce the time needed to charge vehicles, and minimize their downtime. Furthermore, solving the EVCRP can have a significant impact on the transportation industry, as it can lead to the development of more efficient and sustainable transportation systems. By classical means, this problem has been studied by using hybrid genetic algorithms~\cite{liu2022hybrid}, CPLEX solvers~\cite{zhou2021electric}, and column generation algorithms~\cite{liang2021electric}. Some variants have been studied with hybrid combinations of annealing and linear programming~\cite{zaidi2023electric}. 

For solving the EVCRP, we focus on exploiting a novel hybrid quantum--classical search space reduction heuristic method~\cite{Mikel2022a} for imposing constraints in the solutions sampled by quantum algorithms. In particular, we design a Grover based algorithm for sampling partial solutions and we design a classical heuristic for obtaining an approximate global solution. This allows us to greatly reduce the search space of the quantum algorithm so that it can be implemented on currently available NISQ hardware, finding solutions with an approximation ratio close to one.

The rest of the paper is structured as follows: in Section~\ref{sec:description}, we define the classical optimization problem and explain in detail the mapping to the Hamiltonian used in the quantum algorithm; in Section~\ref{sec:SSRheuristic}, we first define our heuristic method for constraint handling and then, in Section~\ref{sec:toyproblem}, we show some preliminary results for a small problem instance. Section~\ref{sec:conclusion} concludes the text with a summary of the article and a perspective of the future work.

\section{Description of the EVCRP problem}\label{sec:description}
The problem we aim at solving in this work is the EVCRP. Previous formulations of the problem involved too many variable for it to be tractable with quantum resources \cite{Jinil2027, Mao2019}, or were oversimplified, allowing for efficient greedy heuristics \cite{Vujanic2016, deller2022}. In this work, we have designed a problem formulation which condenses all the relevant features of this problem while keeping the number of variables low, $\mathcal{O}(3N)$.

\subsection{Definition of the classical problem}

The problem describes the charging of $N$ electric vehicles over the course of a number $T$ of discrete time steps. At each time step $t\in \{1,...T\}$, each vehicle $n\in \{1,...,N\}$ is characterized by three quantities:
\begin{itemize}
    \item Charge level of the vehicle, $\texttt{CL}_{n,t}$.
    \item Flowing energy from the grid to the vehicle, $\texttt{POW}_{n,t}$.
    \item Position of the vehicle at the end of step $t$, $\texttt{POS}_{n,t}$.
\end{itemize}
A general solution of the problem is described by the string of values 
\begin{multline*}
    z=\left(\texttt{CL}_{1,1},\texttt{POW}_{1,1}, \texttt{POS}_{1,1},\dots,\texttt{CL}_{1,T},\texttt{POW}_{1,T},\right.\\
    \left.\texttt{POS}_{1,T},\dots,\texttt{CL}_{N,T},\texttt{POW}_{N,T}, \texttt{POS}_{N,T}\right),
\end{multline*}
from which one can reconstruct the whole charging process and routing of the $N$ vehicles during $T$ time steps.

The goal is to use these variables in order to minimize the value of a target function that represents the overall cost of the charging process, while making sure that the solutions fulfill a given amount of constraints imposed by external factors. In our case, the target function represents the actual cost (i.e., money spent) of the whole charging process and is described by the function 
\begin{equation}
    C(z)=\sum_{n=1}^{N}\sum_{t=1}^T\texttt{POW}_{n,t}\cdot\begin{cases}
    p_t^c & \text{if } \texttt{POW}_{n,t}>0,\\
    p_t^d & \text{if } \texttt{POW}_{n,t}<0,
    \end{cases}
\end{equation}
where $p_t^c$ is the cost of buying one unit of energy at time $t$, and $p_t^d$ the corresponding cost of selling one unit of energy. 
While this function  takes into account only the possible charge state configurations, the positions and charging levels are used to define the constraints on the problem solution. For this problem, we consider the following constraints:
\begin{enumerate}
    \item[C1)] Initial and final positions of each vehicle are fixed, $\texttt{POS}_{n,0}=(\texttt{POS}_\text{ini})_n$, $\texttt{POS}_{n,T}=(\texttt{POS}_\text{fin})_n$.
    \item[C2)] Initial \texttt{CL} is fixed, and the final \texttt{CL} must be above a threshold value, $\texttt{CL}_{n,0}=(\texttt{CL}_\text{ini})_n$, $\texttt{CL}_{n,T}\geq(\texttt{CL}_\text{fin})_n$.
    \item[C3)] A vehicle can only (dis)charge if it does not move, $\texttt{POW}_{n,t}\neq0 \Rightarrow \texttt{POS}_{n,t-1}=\texttt{POS}_{n,t}$. 
    \item[C4)] The \texttt{CL} of the vehicle must change when it (dis)charges, $\texttt{CL}_{n,t}-\texttt{CL}_{n,t-1}=\texttt{POW}_{n,t}$.
    \item[C5)] The car loses $w(i,j)$ energy while moving from point $i$ to $j$, $\texttt{POS}_{n,t-1}\neq\texttt{POS}_{n,t}\Rightarrow\texttt{CL}_{n,t}=\texttt{CL}_{n,t-1}-w(\texttt{POS}_{n,t-1},\texttt{POS}_{n,t})$.
    \item[C6)] The \texttt{CL} of the vehicle must be between certain levels, $\texttt{CL}_\text{min}\leq \texttt{CL}\leq \texttt{CL}_\text{max}$.
    \item[C7)] At each time step, the sum of the power consumed or supplied by the cars must be within the power grid tolerance, $\texttt{POW}_\text{-lim}<\sum_n \texttt{POW}_{n,t}<\texttt{POW}_\text{+lim}$.
\end{enumerate}
Here, we just limit ourselves at stating the list of constraint, while in the next section we explicitly explain how it is possible to include them in the quantum formulation of the problem.

\subsection{Mapping to Quantum}

We map the problem by defining a Hamiltonian such that $H_c\ket{z}=C(z)\ket{z}$. In this way, the solution of the problem can be found by computing the ground state of $H_c$.
We do that by first defining a set of operators $\{\hat{O}\}$ and an orthonormal basis of states $\{\ket{z}\}$ that describe the variables and configurations defined in the previous section. To be more specific, we define these operators such that they are diagonal on the basis  $\{\ket{z}\}$ and their eigenvalues correspond to the classical variables \texttt{POW}, \texttt{CL} and \texttt{POS}:
\begin{align}
    \texttt{POW}_{n,t}=\alpha\ &\rightarrow\ \ket{\alpha_{(\texttt{POW},n,t)}}\\
    \texttt{CL}_{n,t}=\alpha\ &\rightarrow\ \ket{\alpha_{(\texttt{CL},n,t)}}\\
    \texttt{POS}_{n,t}=\alpha\ &\rightarrow\ \ket{\alpha_{(\texttt{POS},n,t)}}
\end{align}
In general, each operator will have a number of eigenvalues equal to the number of possible classical variables associated to it. The natural mapping to the problem is then a composite qudit space.

In the case we have binary variables, we can promote them to the Pauli basis with the following transformation, $\hat{x}_i\rightarrow (\mathbb{1}+\sigma^z_i)/2$. However, the codification we are employing for the quantum variables is in general not binary, so designing a target Hamiltonian that is implementable can be a difficult task. 
In contrast to the Pauli basis for Hermitian operators for two dimensional Hilbert spaces, for higher dimensional spaces there is no consensus about the preferred basis \cite{Bertlmann2008}. In order to avoid making the description of the Hamiltonian too basis specific, we have decided to express it in terms of Dirac delta operators.

If we call $(d_1,d_2,d_3)$ the number of allowed values of the classical variables (\texttt{CL},\texttt{POW},\texttt{POS}), the Hilbert space described by each of these qudits is $\mathbb{C}^{d_i}$. The total Hilbert space  $\mathcal{H}=\mathbb{C}^{\otimes Nd_1}\otimes\mathbb{C}^{\otimes Nd_2}\otimes\mathbb{C}^{\otimes Nd_3}$ has dimension $(d_1d_2d_3)^{NT}$and the basis states are written in the form
\begin{equation}\ket{z}=\bigotimes_{n=1}^N\bigotimes_{t=1}^T\ket{\alpha_{(\texttt{CL},t,n)}}\otimes\ket{\alpha_{(\texttt{POW},n,t)}}\otimes\ket{\alpha_{(\texttt{POS},n,t)}}.
\end{equation}
The cost Hamiltonian $H_C$ can thus be defined as 
\begin{equation}\label{eq:cost_hamiltonian}
    \small
    H_c=\sum_{n=1}^N\sum_{t=1}^T\left(\sum_{j=-(d_2-1)/2}^{(d_2-1)/2}j\left.\begin{cases}
        p_t^c &\text{if } j>0\\
       p_t^d &\text{if } j<0\\
    \end{cases}\right\}\ket{j_{(\texttt{POW},t,n)}}\bra{\cdot}\right).
\end{equation}
For readability, we employ the simplified notation $\ket{\psi_{(\texttt{A},t)}}\bra{\psi_{(\texttt{A},t)}}\equiv\ket{\psi_{(\texttt{A},t)}}\bra{\cdot}$, where the label $(\texttt{A},t)$ refers to the qudit encoding the information about the variable \texttt{A} for the time step $t$. Constraints C1-C5 can be introduced as energy contributions that lower the energy of the allowed states. Constraint C6 can be introduced by defining a one-to-one mapping such that we limit the dimension of the qudits encoding the information about the $\texttt{CL}$ to be equal to the number of allowed charging levels.

The contributions to the Hamiltonians for each respective constraint are introduced as follows for the different cars:
\begin{align}
    H_1=&-\ket{\left(\texttt{POS}_{\text{ini}}\right)_{(\texttt{POS},0)}}\bra{\cdot}-\ket{(\texttt{POS}_\text{fin})_{(\texttt{POS},T)}}\bra{\cdot}\,,\label{eq:const1}\\
    H_2=&-\ket{(\texttt{CL}_\text{ini})_{(\texttt{CL},0)}}\bra{\cdot}-\sum_{j=\texttt{CL}_\text{fin}}^{\texttt{CL}_\text{max}}\ket{j_{(\texttt{CL},T)}}\bra{\cdot}\,,\\
    H_3=&-\sum_{i\neq j}^{\#\texttt{POS}}\ket{i_{(\texttt{POS},t-1)};j_{(\texttt{POS},t)};0_{(\texttt{POW},t)}}\bra{\cdot}\,,\\
    H_4=&-\sum_{t=1}^T\sum_{j=\texttt{CL}_\text{min}}^{\texttt{CL}_\text{max}}\sum_{\substack{k=-\texttt{POW}_\text{max}\\k\neq0}}^{\texttt{POW}_\text{max}}\nonumber\\
        &\ket{j_{(\texttt{CL},t-1)};\left(j+k\right)_{(\texttt{CL},t)};k_{(\texttt{POW};t)}}\bra{\cdot}\,,
\end{align}
\begin{multline}\label{eq:const5}
    H_5=-\sum_{t=1}^T\sum_{i=\texttt{CL}_\text{min}}^{\texttt{CL}_\text{max}}\sum_{j\neq k}^{\#\texttt{POS}}\\
        \left|i_{(\texttt{CL},t-1)};\left(i-w(j,k)\right)_{(\texttt{CL},t)};j_{(\texttt{POS},t-1)};k_{(\texttt{POS},t)}\right>\bra{\cdot}\,.
\end{multline}
In $H_4$, if $j+k$ is outside the encoded values for the \texttt{CL}, then we discard the term, 
and similarly in $H_5$ if $i-w(j,k)$ is not encoded. We then construct the full Hamiltonian by summing each term
\begin{equation}\label{eq:full_hamiltonian}
    H=H_c+\sum_{i=1}^{5} \lambda_iH_i
\end{equation}
and look for the ground state. Here, the parameters $\lambda_i>0$ are factors that lower the energy of the feasible states to ensure that the ground state corresponds to a feasible solution.  

\section{Search Space Reduction heuristic}\label{sec:SSRheuristic}
A seen in the previous section, if we map the full optimization problem to the problem of finding the ground state of the Hamiltonian in Equation \ref{eq:full_hamiltonian}, we will need a large amount of resources. In particular, we will need a number of qudits that scales exponentially with the number of time steps and the number of vehicles. This scaling renders the problem unsolvable for the current NISQ-era devices.

For solving the EVCRP, we will employ a Search Space Reduction heuristic. This heuristic aims at employing the limited quantum resources in a more efficient way, by dividing constrained optimization problems into a two step problem. The main idea is to classify the constraints into two types: global and partial constraints. On one hand, partial constraints act in parallel on a small number of variables. On the other hand, global constraints have to be enforced on a large number of variables at the same time. It follows, that in order to solve the problem, the full solution must fulfill all the partial constraints. In other words, full solutions can be built as a concatenation of feasible partial constraints. This distinction allows us to target the complete problem in two steps. First, we need to find the set of feasible partial solutions. Second, we can try to find the optimum for the problem, or at least a good approximation, by selecting a combination of feasible partial solution that qualifies as a full solution. This kind of strategy has been employed previously for solving the Bin Packing Problem \cite{Mikel2022a, Mikel2022b}.

The main advantage that we obtain from employing this strategy is that we can benefit from the advantages of quantum computing on problems whose scale is still manageable in the NISQ era. The noise levels on the current quantum devices, the small amount of qubits and their limited connectivity, prevent us from implementing long quantum circuits tackling medimum-large scale problems. This is why this hybrid heuristic may offer one solution to this limitation, as it employs the quantum resources only for the problem of finding feasible partial constraints. After that, a classical procedure generates the solution to the full problem. 

\subsection{Partial problem sampling}

In the EVCRP, the partial/global constraint distinction can be made in a natural way. The constraints acting on the different cars (i.e., constraints C1 to C6) can be labeled as the partial constraints. Then, global solutions can be constructed as a combination of one of the partial solutions from each of the vehicles. For the task of sampling these partial solutions, we review here three possible algorithms.

\subsubsection{Ground state search}
As stated in the previous section, classical problems can be mapped to the problem of finding the ground state of a Hamiltonian. In this case, our objective is to sample all feasible partial solutions. Since the information about the cost function is irrelevant for the feasibility of the solutions, the Hamiltonian we employ is the sum of the Hamiltonian terms encoding each of the constraints, this is, the $H_1$ to $H_5$ terms form Equations~\ref{eq:const1}-\ref{eq:const5}. This Hamiltonian can be then employed as the target Hamiltonian for various algorithms, whose goal is to search for its ground state. The most well known algorithms for doing so are quantum annealing and QAOA, which both have been extensively studied in the literature.

\subsubsection{Grover}\label{sec:grover}

Even though it seems counterintuitive to employ the Grover algorithm for solving an optimization problem, for the search space reduction heuristic we only need to find feasible partial solutions. Since the answer if a partial solution is valid is only yes or no, we can generate an oracle for this task. The generation of a Grover oracle in general is a complex task, and automatic compilers tend to provide suboptimal circuits~\cite{Nikolaeva2023, Seidel2023}. In this case, it is even more challenging, since we are working with $k$-dimensional qudits. As the task of giving an efficient circuit for implementing the sampling algorithm is out of the scope of this work, we will employ the analytic results from~\cite{Boyer1998} for analyzing its performance. The success probability of the Grover algorithm for finding one of $K$ solutions in a $N$ dimensional Hilbert space is
\begin{equation}
    P(N,K,i)=\sin\left((2i+1)\arcsin\left(\sqrt{\frac{K}{N}}\right)\right),
\end{equation}
where $i$ is the number of Grover iterations.

\begin{figure}[t!]
    \centering
    \includegraphics[width=\linewidth,trim={50pt 0pt 50pt 30pt},clip]{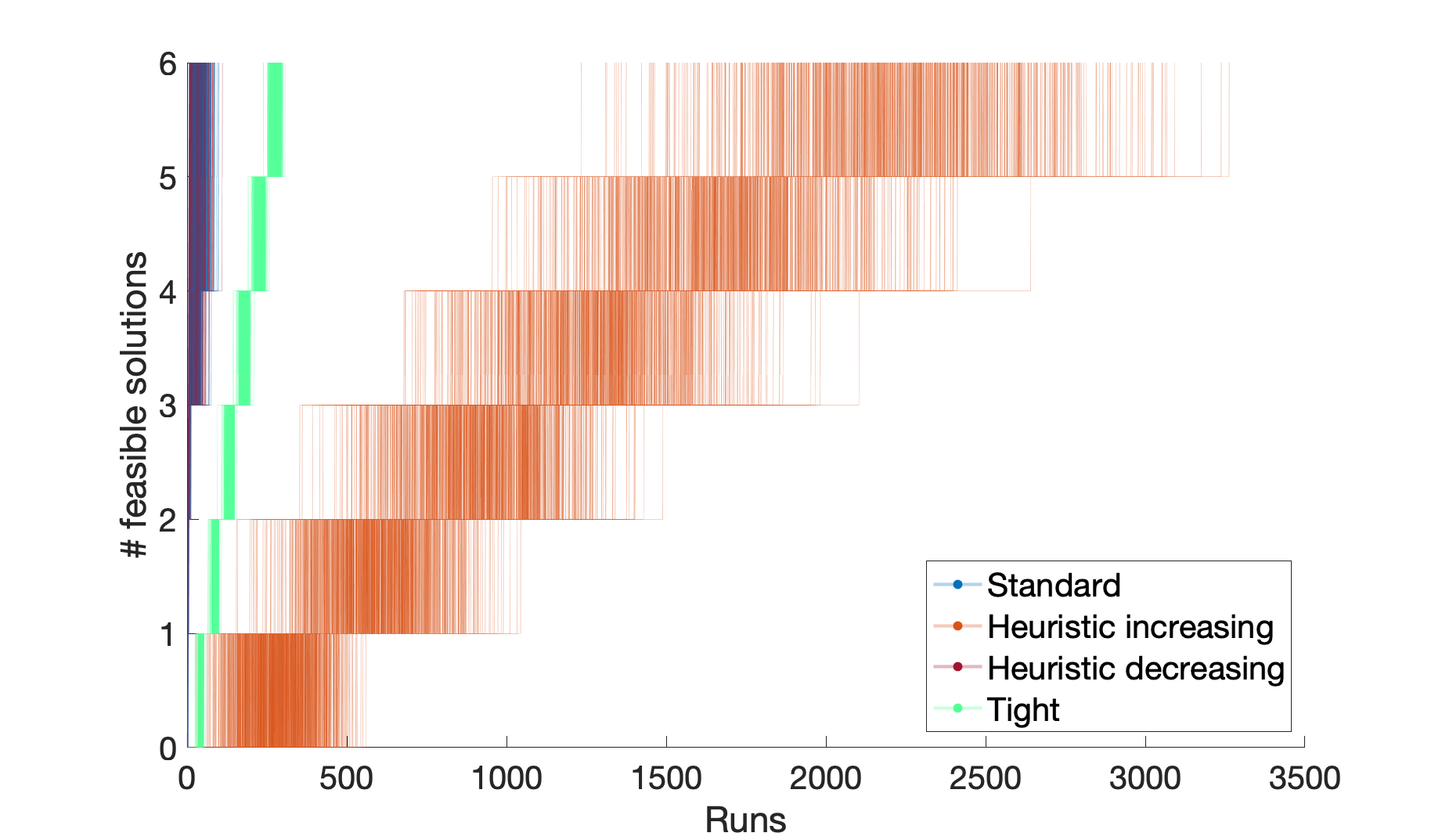}
    \includegraphics[width=\linewidth,trim={50pt 0pt 50pt 30pt},clip]{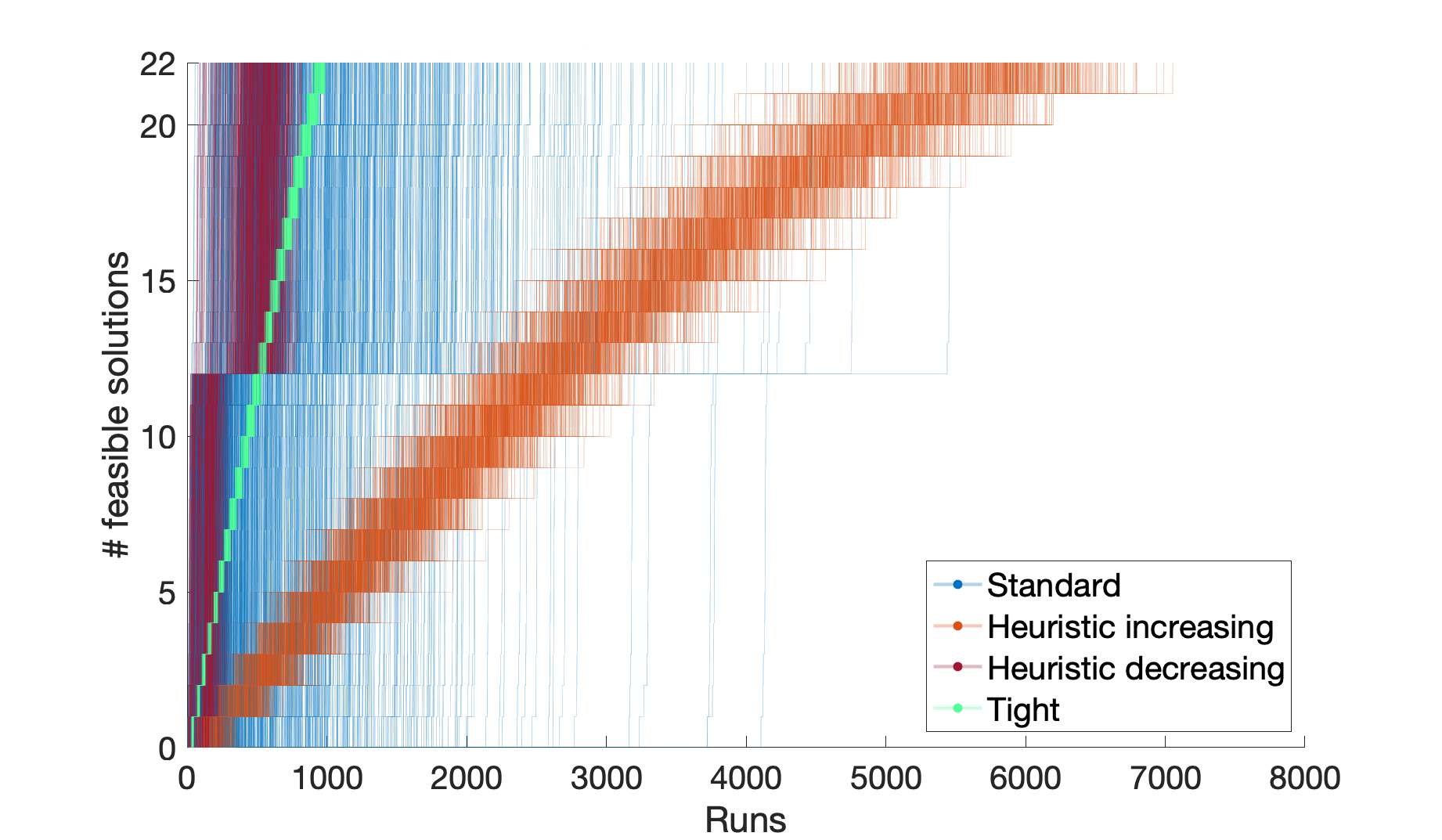}
    \includegraphics[width=\linewidth,trim={50pt 0pt 50pt 30pt},clip]{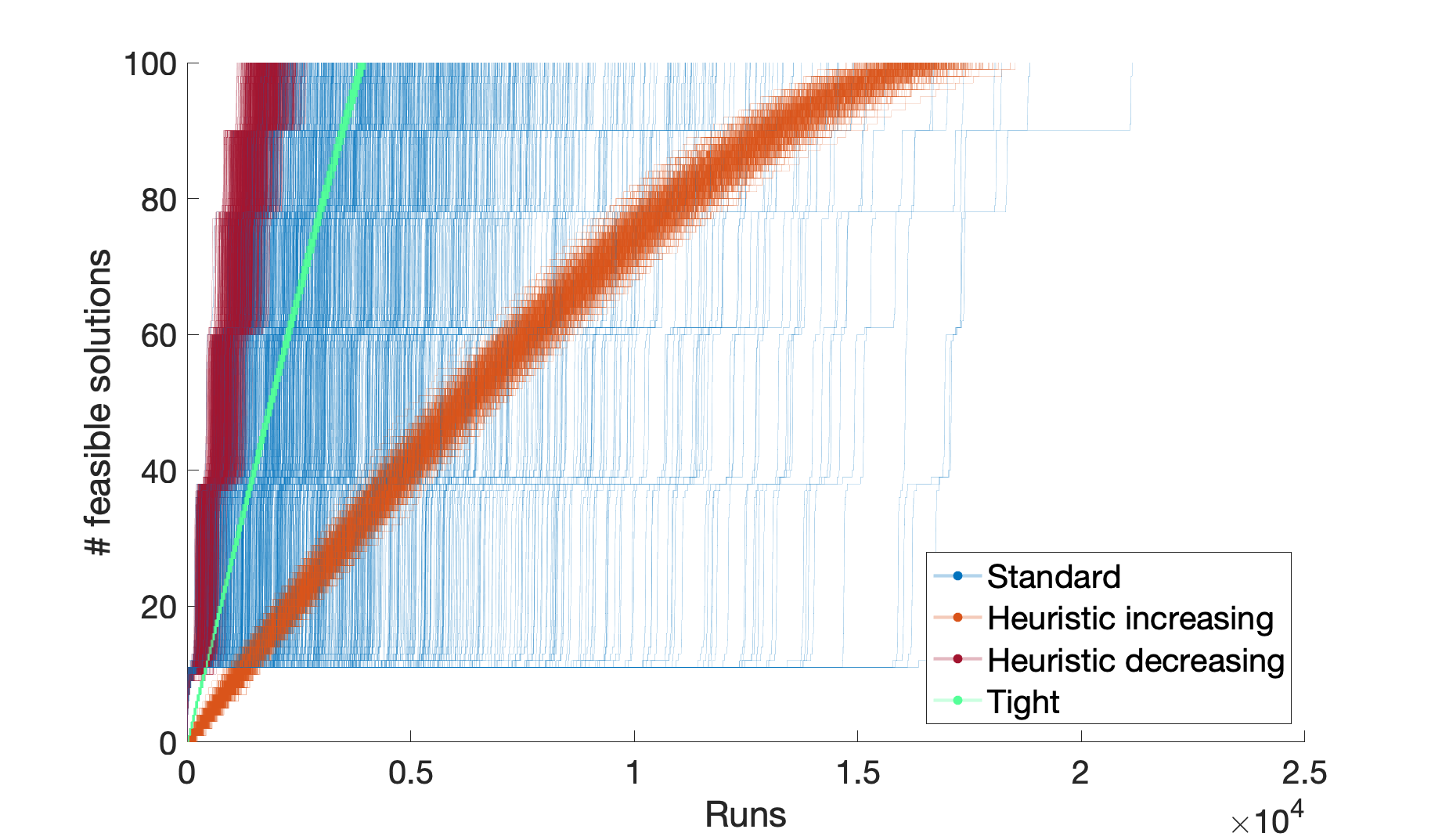}
    \caption{Number of sampled partial solutions with a Grover algorithm in a $N\sim2\cdot10^8$ dimensional Hilbert space vs number of circuit runs. From top to bottom, we aim at obtaining $K=6,22,100$ feasible solutions. We compare 4 different strategies for selecting the number of Grover iterations for each run. We conclude that the heuristic in which we increase the number of iterations in each consecutive run outperforms other tested strategies for the sampling task.}
    \label{fig:grover}
\end{figure}

In \cite{Boyer1998}, a strategy was proposed for finding one solution without prior knowledge about the number of solutions. Our problem is slightly different, since our objective is to sample as many of these solutions as possible. This means, that each time we obtain a new solution, the size of the target set decreases by one. For making this sampling more efficiently, we propose a new heuristic for selecting the number of j iterations in each run. The new strategy consists in a deterministic sweep from $\sqrt{N}$ iterations to 1 (heuristic decreasing), and vice versa (heuristic increasing). When hitting the limit, we reset the number of iterations to the initial position. In Figure~\ref{fig:grover}, we see that the decreasing strategy outperforms the rest of the strategies for a relatively small number of feasible solutions. In these cases, is also worth noting that running the algorithm with a constant number of iterations, $\sqrt{N}$, also performs well. However, as we increase the number of feasible solutions, its performance decreases rapidly.

\subsubsection{Classical sampling}

As for other satisfiability problems, we have a plethora of classical algorithms for sampling feasible partial solutions. However, since the constraints generate a rich structure for the problem, we can always find a way of employing prior information to speedup the sampling. To be most efficient, this requires a deep understanding of the problem and its symmetries. Here, instead of trying to find an efficient classical method, we employ a naive random sampling method for setting the baseline for a comparison.

\subsection{Global solution search}

The second step of the proposed heuristic is to generate full solutions to the problem and optimize them according to the cost function. Since the full solutions involve a high number of variables, we decide to solve this problem employing only classical resources. For the EVCRP the goal is to select one feasible partial solution for each of the cars. We propose two different strategies for this task.

\subsubsection{Bruteforce}

Assuming that we have correctly sampled all the partial solutions, we can obtain the optimal solution to the problem by trying all possible combinations of them. For small problems, this strategy is affordable. However, as the number of solutions per car or the number of cars grow, the number of combinations increases rapidly. For this reason, it is essential to find a heuristic that can obtain a good approximate solution employing as few resources as possible.

\subsubsection{Greedy tree}

The cost function for this problem allows us to compute it as the sum of the cost of each vehicle. This allow us to order the partial solutions in terms of their cost. Let us order these partial solutions in ascending order, assigning the index 0 to the one with lowest cost. If we ignore the global constraints, the best possible overall solution would be the one that takes the best solution for each of the vehicles. In this case, the sum of the indices of the solutions from which we have built the solution would be 0. However, when imposing the global constraints, it is unlikely that this naive solution is allowed. We propose a tree search strategy where we explore the solutions by levels, each of them corresponding to the sum of the indices of the solutions. Even though this strategy cannot assure to output the optimal solution for every problem, it is likely to be close to the optimal one. 

\section{Toy problem}\label{sec:toyproblem}

For testing that the proposed heuristic is a valid approach for solving the EVCRP, we have benchmarked it employing a synthetic toy problem. 
At this stage, all the testing has been done analytically or by numerical calculations. 

\begin{figure}[t]
    \centering
    \includegraphics[width=0.8\linewidth,trim={0 20pt 0 15pt},clip]{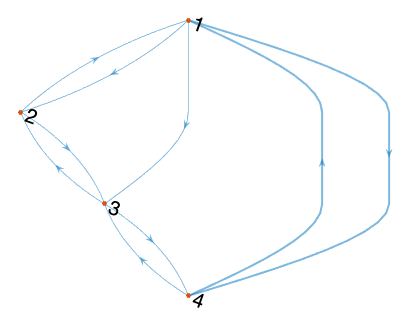}
    \caption{Graph for the routes between the 4 different nodes for the toy problem, with $w=[0, 0, 0, 1;0, 0, 0, \infty;\infty, 0, 0, 0;1, \infty, 0, 0]$. The thicker lines between nodes 1 and 4 marks routes which have a cost of 1 unit of energy for traversing them. The rest of the thin lines depict routes where the energy cost is 0. All the edges can be traversed in a single time step.}
    \label{fig:map}
\end{figure}

The problem instance tries to optimize the charging and routing schedule of $N=4$ cars in $T=4$ time steps. The graph where the vehicles move has 4 nodes (\texttt{POS}$=\{1,2,3,4\}$), with some missing connections and a road that requires the expenditure of energy to traverse it, as shown in Figure \ref{fig:map}. The initial/final charge and position requirements are shown in Table \ref{tab:constraints}. For the power grid requirements for the global constraint, we set the maximum total power to $\texttt{POW}_{\pm\text{lim}}=\pm3$. The energy market is defined by the energy buying price, $p^c=[3, 5, 4, 5]$, and the selling price, $p^d=[2, 4.5, 3.5, 4]$. Vehicles can (dis)charge one energy unit per turn \texttt{POW}$=\{-1,0,1\}$ and we set \texttt{CL}$=\{1,2,3,4,5\}$.

\begin{table}[h]
\caption{Initial and final requirements of the position and the charge level of the different vehicles for the toy problem.}
\label{tab:constraints}
\centering
\normalsize
\begin{tabular}{c|c|c|c|c|}
\cline{2-5}
                                & \texttt{POS}$_{\text{ini}}$ & \texttt{POS}$_{\text{fin}}$ & \texttt{CL}$_{\text{ini}}$ & \texttt{CL}$_{\text{fin}}$ \\ \hline
\multicolumn{1}{|c|}{Vehicle 1} & 2                & 4                & 3                & 5                \\ \hline
\multicolumn{1}{|c|}{Vehicle 2} & 1                & 3                & 1                & 3                \\ \hline
\multicolumn{1}{|c|}{Vehicle 3} & 2                & 3                & 1                & 4                \\ \hline
\multicolumn{1}{|c|}{Vehicle 4} & 4                & 1                & 3                & 4                \\ \hline
\end{tabular}
\end{table}

\subsection{Results}

We have run numerical simulations for both of the sampling of feasible partial solutions and the generation of full solutions from the partial ones.

For the first task, we have run a numerical calculation for the Grover algorithm. For this, we employed the four different strategies mentioned in Section \ref{sec:grover} for selecting the number of Grover iterations in each run. Since the goal of this numerical test was to test the strategies without any prior knowledge, the program does not have any access to the number of feasible solutions for the problem. The results are similar to the ones shown in Figure~\ref{fig:grover}, in which for the vehicles with a low number of feasible partial solutions the standard and the decreasing heuristics are equally the best performing ones. As the number of feasible partial solutions increases, the performsnce of the standard strategy decreases. In all cases, the decreasing heuristic displays the best performance for the task of sampling, regardless of the number of feasible solutions.

For the second subroutine of our heuristic, we tested the greedy tree search strategy for this problem, and we compared it to the optimal solution. For this, we first assumed that the sampling strategy is able to extract all the feasible partial solutions. The optimal solution is computed with a brute-force algorithm, which exhausts all combinations of partial solutions. For this instance, the solutions range from a cost of 31 to 48 in jumps of 1. After just a single step, our heuristic outputs a solution with a cost of 36, which corresponds to an approximation ratio 1.16, defined as the ratio between the obtained cost value and the optimal one, $\varepsilon=f/f_\text{opt}$. This result shows that this strategy is able to find competitive results with a reduced number of steps. We are confident that this strategy will successfully generate solutions for instances where a brute-force approach is not an option due to the high number of combinations.

The performance of the search space reduction can not be conclusively addressed by the success for a single problem instance. However, in this case, the reduction in the search space can be directly seen. For the toy problem, the original search space has a size of order $\sim10^{28}$. The encoding we have selected for the sampling of partial solutions generates a Hilbert space of dimension $\sim10^8$. The number of feasible partial solutions is $\{6,22,4,19\}$ for each respective vehicle. The combination of these partial solutions generates a space of $\sim10^4$ different global solutions, from which only $\sim300$ are allowed by the global constraint. This reduction in the search space can be advantageous for tackling real-size EVCRPs by jointly employing quantum and classical resources more efficiently. 

\section{Conclusion}\label{sec:conclusion}
In this work, we proposed an enriched version of the Electric Vehicle charging problem which includes discharging and routing of the vehicles. This new formulation enlarges the space of solutions and introduces new constraints into the problem. A useful characteristic of this new constraint set is that it allows us to classify the constraints as local or global, thus permitting us to employ a search space reduction heuristic. We showed how to encode the partial constraints into qudit quantum architectures. In particular, we proposed a sampling of the partial solutions by mapping the problem into a ground state problem. We also proposed a Grover algorithm, for which we tested its performance in a noiseless setup. Then, we obtained the global solution to the problem through a greedy heuristic. We tested this strategy on a toy problem, for which we obtain a global solution with an approximation ratio of 1.16. From here, we conclude that the proposed heuristic may be a valid strategy for solving combinatorial optimization problems.

For a future line of research, the next steps should consider a practical implementation of the quantum subroutine on a quantum device, in particular also harnessing the ability to manipulate qudits that has been demonstrated or proposed for various platforms \cite{Ringbauer2022,Low2020Senko,Kasper2022,Blok2021Qutrit}. Even though in this work we have focused on a Grover algorithm for the sampling, future works should also consider variational quantum circuits and adiabatic processes. There is also a need for developing a proper framework for the search space reduction heuristic, which is still in development. We hope that the techniques described here can in the future help solving realistic transport and logistic problems by employing quantum resources efficiently.

\section*{Acknowledgements}

PH and SS acknowledge funding by the European Union under Horizon Europe Programme, Grant Agreement 101080086 (NeQST). Views and opinions expressed are however those of the author(s) only and do not necessarily reflect those of the European Union or European Climate, Infrastructure and Environment Executive Agency (CINEA). Neither the European Union nor the granting authority can be held responsible for them. MGdA and MS acknowledge support from EU FET Open project EPIQUS (899368), HORIZON-CL4-2022-QUANTUM01-SGA project 101113946 OpenSuperQPlus100 of the EU Flagship on Quantum Technologies, Spanish Ram\'on y Cajal Grant RYC-2020-030503-I and project Grant No. PID2021-125823NA-I00 funded by MCIN/AEI/10.13039/501100011033 “ERDF A way of making Europe”, “ERDF Invest in your Future”, and IKUR Strategy under the collaboration agreement between Ikerbasque Foundation and BCAM on behalf of the Department of Education of the Basque Government. MGdA acknowledges support from the UPV/EHU and TECNALIA 2021 PIF contract call. 
AB acknowledges funding from the Honda Research Institute Europe GmbH.
PH acknowledges fruitful discussions and collaborations within the project "EnerQuant" of the Bundesministerium für Wirtschaft und Energie (project ID 03EI1025C) and Q@TN — Quantum Science and Technology in Trento, the joint laboratory between University of Trento, FBK-Fondazione Bruno Kessler, INFN-National Institute for Nuclear Physics, and CNR-National Research Council, and acknowledges funding from Provincia Autonoma di Trento and the ERC Starting Grant StrEnQTh (project ID 804305). 
\bibliographystyle{IEEEtran}    
\bibliography{main.bib}

\end{document}